\documentstyle[twoside,fleqn,espcrc2]{article}
\input{epsfig.sty}

\newcommand{\AmS}{{\protect\the\textfont2
  A\kern-.1667em\lower.5ex\hbox{M}\kern-.125emS}}
\def\beq{\begin{equation}}
\def\eeq{\end{equation}}
\def\beqa{\begin{eqnarray}}
\def\eeqa{\end{eqnarray}}

\title{A model for $J/\psi$ - kaon cross section}
\author{R.S. Azevedo and M. Nielsen
\address{Instituto de F\'{\i}sica, 
        Universidade de S\~{a}o Paulo, \\
        C.P. 66318,  05389-970 S\~{a}o Paulo, SP, Brazil}}

\begin{document}

\begin{abstract}
We calculate the cross section for the dissociation of $J/\psi$ by kaons
within the framework of a meson exchange model. We find that, depending
on the values of the coupling constants used, the cross section can vary 
from 5 mb to 30 mb at $\sqrt{s}\sim5$ GeV.
\end{abstract}

\maketitle

In  relativistic  heavy  ion  collisions $J/\psi$ suppression has
been recognized as an important tool  to  identify  the  possible
phase  transition  to  quark-gluon  plasma (QGP) \cite{ma86} (for a 
review of data  and interpretations see refs.~\cite{vo99,ge99}). 
Since there is no direct experimental information on $J/\psi$ absorption cross
sections by hadrons, several theoretical approaches have been proposed to
estimate their values.
In order to elaborate a theoretical description of the phenomenon, we have
first to choose the relevant degrees of freedom. Some
approaches were based on charm quark-antiquark dipoles interacting with the
gluons of a larger
(hadron target) dipole \cite{bhp,kha2,lo} or quark exchange between two
(hadronic) bags \cite{wongs,mbq}, or QCD sum rules \cite{nnmk,dlnn,dklnn}, 
whereas other works used the meson exchange mechanism
\cite{mamu98,haglin,linko,haga,osl,nnr}.
In this case it is not easy to decide
in favor of quarks or hadrons because we are dealing with charm quark bound
states, which are small and massive enough to make perturbation theory
meaningful, but not small enough to make
non-perturbative effects negligible \cite{nnmk,dlnn,dklnn,dnn}.

The meson exchange approach was applied basically 
to $J/\psi-\pi$ and $J/\psi-\rho$ cross sections, with the only exception
of ref.~\cite{haglin} where $J/\psi-K$ cross section was also estimated.
However, as pointed out in ref.~\cite{linko}, there are some inconsistencies
in the Lagrangians defined in ref.~\cite{haglin}. In this work we will
evaluate the $J/\psi-K$ cross section using a meson-exchange model
as in ref.~\cite{haglin}, but we will treat the VVV and four-point couplings
in the effective Lagrangians as in ref.~\cite{linko}.

As in refs.~\cite{haglin,linko,haga,osl} we start with the SU(4) Lagrangian
for the pseudo-scalar and vector mesons. The effective Lagrangians 
relevant for the study of the $J/\psi$ absorption by kaons are:
\beq
{\cal L}_{K DD_s^*}=ig_{K DD_s^*}~ 
D_s^{* \mu}\left ( \bar D \partial_\mu \bar K - 
(\partial_\mu \bar D) \bar K \right ) + {\rm H.c.}~ , \label{kdds} 
\eeq
\beq
{\cal L}_{K D_sD^*}=ig_{K D_sD^*}~ 
D^{* \mu} \left ( \bar D_s \partial_\mu K - 
(\partial_\mu \bar D_s) K \right ) + {\rm H.c.}~ , \label{kdsd}
\eeq
\beqa
{\cal L}_{\psi DD}&=&ig_{\psi DD}~ \psi^\mu 
\left ( D \partial_\mu \bar D -(\partial_\mu D) \bar D \right ) ~ ,
\label{jdd}\\ 
{\cal L}_{\psi D_sD_s}&=&ig_{\psi D_sD_s}~ \psi^\mu 
\left ( D_s \partial_\mu \bar D_s -(\partial_\mu D_s) \bar D_s \right ) ~ ,
\label{jdsds} \\
{\cal L}_{\psi D^*D^*}&=& ig_{\psi D^*D^*}~
\left [ \psi^\mu \left ( (\partial_\mu D^{* \nu}) \bar {D^*_\nu} 
- D^{* \nu} \partial_\mu \bar {D^*_\nu} \right )\right.\nonumber \\
&+&\left ( (\partial_\mu \psi^\nu) D^*_\nu -\psi^\nu 
\partial_\mu D^*_\nu 
\right )\bar {D^{* \mu}}\nonumber \\ 
&+&\left. D^{* \mu} \left ( \psi^\nu \partial_\mu \bar {D^*_\nu} -
(\partial_\mu \psi^\nu) \bar {D^*_\nu} \right ) \right ] ~ , \label{jdede}\\ 
{\cal L}_{\psi D_s^*D_s^*}&=& ig_{\psi D_s^*D_s^*}~
\left [ \psi^\mu \left ( (\partial_\mu D_s^{* \nu}) {\bar D}_{s\nu}^* 
- D_s^{* \nu} \partial_\mu {\bar D}_{s\nu}^* \right )\right.\nonumber \\
&+&\left ( (\partial_\mu \psi^\nu) D_{s\nu}^* -\psi^\nu 
\partial_\mu D_{s\nu}^* 
\right ){\bar D}_s^{* \mu} \nonumber \\ 
&+&\left . D_s^{* \mu} \left ( \psi^\nu \partial_\mu {\bar D}_{s\nu}^* -
(\partial_\mu \psi^\nu) {\bar D}_{s\nu}^* \right ) \right ] ~ , 
\label{jdesdes}\\ 
{\cal L}_{K \psi D_sD^*}&=&-g_{K \psi D_sD^*}~
\psi^\mu \left ( D^*_\mu K \bar D_s
+ D_s \bar K \bar {D^*_\mu} \right )  ~ , \label{kjdsd} \\ 
{\cal L}_{K \psi DD_s^*}&=&-g_{K \psi DD_s^*}~
\psi^\mu \left ( {\bar D}^*_{s\mu} K  D
+  \bar D \bar K  D^*_{s\mu} \right )  ~ , \label{kjdds}
\end{eqnarray}
where we have defined the charm meson and kaon iso-doublets $D\equiv(D^0,D^+)$,
$D^*\equiv(D^{*0},D^{*+})$ and $K\equiv(K^0,K^+)$. 

The processes we want to study for the absorption of $J/\psi$ by kaons are:
\beqa
K J/\psi \rightarrow D_s {\bar D}^*, && \bar K J/\psi \rightarrow D^* 
{\bar D}_s
\label{pro1}\\ 
K J/\psi \rightarrow D_s^* \bar D, && \bar K J/\psi \rightarrow D {\bar D}_s^*.
\label{pro2}
\eeqa
The two processes in eqs.(\ref{pro1}) and (\ref{pro2}) have the same 
cross section.
Therefore, in Fig.~1 we only show the diagrams for the first process
in eqs.~(\ref{pro1}) and (\ref{pro2}).

Defining  the four-momentum of the kaon, 
the $J/\psi$, the vector and pseudo-scalar final-state mesons
respectively by $p_1$, $p_2$, $p_3$ and $p_4$, the full amplitude for the 
first process $K \psi \rightarrow D_s \bar D^*$, 
without isospin factors and before summing and averaging over external spins, 
is given by
\begin{figure} \label{fig1}
\centerline{\epsfig{figure=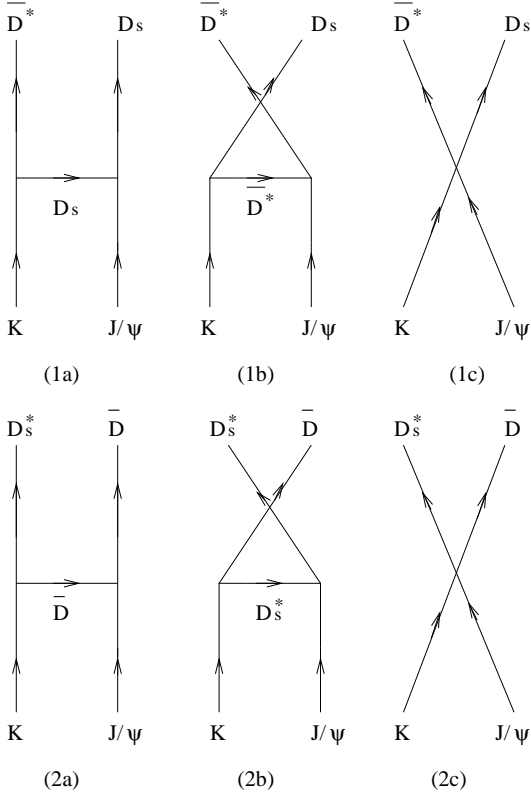,width=7cm}}
\vspace{-1cm}
\caption{\small{Diagrams for $J/\psi$ absorption processes:   
1) $K \psi \rightarrow D_s {\bar D}^*$;
2) $K \psi \rightarrow D_s^* \bar D$. Diagrams for
the processes $\bar K \psi \rightarrow {\bar D}_s D^*$ and
$\bar K \psi \rightarrow {\bar D}_s^* D$
are similar to (1a)-(1c) and (2a)-(2c) respectively, but with each particle 
replaced by its anti-particle.}}  
\end{figure}

\begin{eqnarray}
{\cal M}_1 \equiv {\cal M}_1^{\nu \lambda} 
~\epsilon_{2 \nu} \epsilon_{3 \lambda} 
=\left ( \sum_{i=a,b,c} {\cal M}_{1 i}^{\nu \lambda} \right )
\epsilon_{2 \nu} \epsilon_{3 \lambda}, 
\label{m1}
\end{eqnarray}
with
\begin{eqnarray}
{\cal M}_{1a}^{\nu \lambda}&=& -g_{K D_s D^*} g_{\psi D_s D_s}~
(-2p_1+p_3)^\lambda \left (\frac{1}{t-m_{D_s}^2} \right ) \nonumber \\
&\times&(p_1-p_3+p_4)^\nu, \nonumber \\
{\cal M}_{1b}^{\nu \lambda}&=& g_{K D_s D^*} g_{\psi D^* D^*}~
(-p_1-p_4)^\alpha \left ( \frac{1}{u-m_{D^*}^2} \right )\nonumber \\
&\times&
\left [ g_{\alpha \beta}-\frac{(p_1-p_4)_\alpha (p_1-p_4)_\beta}{m_{D^*}^2}
\right ] 
\left [ (-p_2-p_3)^\beta g^{\nu \lambda}\right.\nonumber \\
&+&\left.(-p_1+p_2+p_4)^\lambda g^{\beta \nu}
+(p_1+p_3-p_4)^\nu g^{\beta \lambda} \right ] , \nonumber \\
{\cal M}_{1c}^{\nu \lambda}&=& -g_{K \psi D_s D^*}~ g^{\nu \lambda},
\label{pij1}
\end{eqnarray}
where $t=(p_1-p_3)^2$ and $u=(p_1-p_4)^2$.

Similarly, the full amplitude for the second process 
$K \psi \rightarrow D_s^* \bar D$ is given by
\begin{eqnarray}
{\cal M}_2 \equiv {\cal M}_2^{\nu \lambda} 
~\epsilon_{2 \nu} \epsilon_{3 \lambda} 
=\left ( \sum_{i=a,b,c} {\cal M}_{2 i}^{\nu \lambda} \right )
\epsilon_{2 \nu} \epsilon_{3 \lambda}, 
\label{m2}
\end{eqnarray}
with
\begin{eqnarray}
{\cal M}_{2a}^{\nu \lambda}&=& -g_{K D D_s^*} g_{\psi D D}~
(-2p_1+p_3)^\lambda \left (\frac{1}{t-m_{D}^2} \right ) \nonumber \\
&\times&(p_1-p_3+p_4)^\nu, \nonumber \\
{\cal M}_{2b}^{\nu \lambda}&=& g_{K D D_s^*} g_{\psi D_s^* D_s^*}~
(-p_1-p_4)^\alpha \left ( \frac{1}{u-m_{D_s^*}^2} \right )\nonumber \\
&\times&
\left [ g_{\alpha \beta}-\frac{(p_1-p_4)_\alpha (p_1-p_4)_\beta}{m_{D_s^*}^2}
\right ] 
\left [ (-p_2-p_3)^\beta g^{\nu \lambda}\right.\nonumber \\
&+&\left.(-p_1+p_2+p_4)^\lambda g^{\beta \nu}
+(p_1+p_3-p_4)^\nu g^{\beta \lambda} \right ] , \nonumber \\
{\cal M}_{2c}^{\nu \lambda}&=& -g_{K \psi D D_s^*}~ g^{\nu \lambda}.
\label{pij2}
\end{eqnarray}

We can see that the differences between these two processes are basically due
to the meson exchanged.
It can be shown \cite{linko} that the full amplitudes 
${\cal M}_{i}^{\nu \lambda}$ (for $i=1,2$) given above satisfy current 
conservation: ${\cal M}_{i}^{\nu \lambda}p_{2\nu}={\cal M}_{i}^{\nu \lambda}
p_{3\lambda}=0$.

After averaging (summing) over initial (final) spins 
and including isospin factors, the cross sections 
for these two processes are given by 
\beqa\label{jpion}
\frac {d\sigma_i}{dt}&=& \frac {1}{192 \pi s p_{0,\rm cm}^2} 
{\cal M}_i^{\nu \lambda} {\cal M}_i^{*\nu^\prime \lambda^\prime}
\left ( g_{\nu \nu^\prime}-\frac{p_{2 \nu} p_{2 \nu^\prime}} {m_2^2} \right )
\nonumber\\
&\times&\left ( g_{\lambda \lambda^\prime}
-\frac{p_{3 \lambda} p_{3 \lambda^\prime}} {m_3^2} \right ),
\end{eqnarray}
with $s=(p_1+p_2)^2$, and  
\begin{eqnarray}
p_{0,\rm cm}^2=\frac {\left [ s-(m_1+m_2)^2 \right ]
\left [ s-(m_1-m_2)^2 \right ]}{4s}
\end{eqnarray}
is the squared three-momentum of initial-state mesons in the 
center-of-momentum (c.m.) frame.

To estimate the cross sections we have first to determine the coupling 
constants of our effective Lagrangians.
Exact SU(4) symmetry 
would give the following relations among the coupling constants:
\begin{eqnarray}
&&g_{K D_sD^*}=g_{K DD_s^*}=\frac{g}{2\sqrt{2}}~, \nonumber \\
&&g_{\psi DD}=g_{\psi D_sD_s}=g_{\psi D^* D^*}=g_{\psi D_s^* D_s^*}=
\frac{g}{\sqrt 6}~, \nonumber \\
&&g_{K \psi D_sD^*}=g_{K \psi DD_s^*}=\frac{g^2}{4 \sqrt 3}~.
\label{su4}
\end{eqnarray}

\begin{figure}[htb]
\centerline{\epsfig{figure=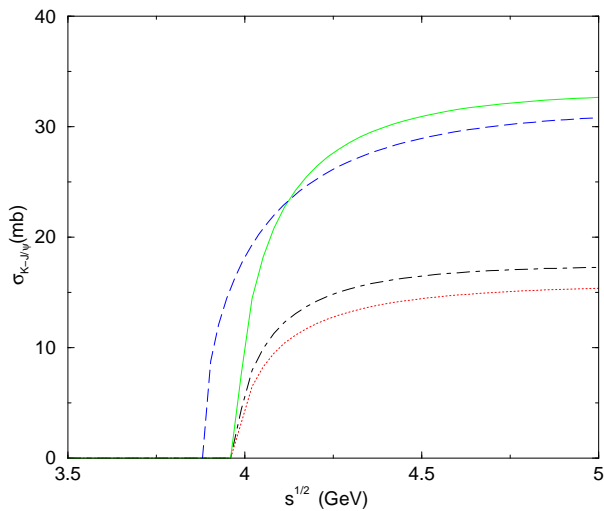,width=8cm}}
\vspace{-1cm}
\caption{\small{Total cross sections of the processes $J/\psi~\mbox{kaons}
\rightarrow$ $\bar{D}^*~D_s + D^*~\bar{D}_s$ (dot-dashed line), $\bar{D}~
D_s^* +\bar{D}_s^*~D$ (dotted line). 
The solid line gives the total $J/\psi$ dissociation by kaons cross section. 
For comparison, the dashed line gives the $J/\psi~\pi\rightarrow$ 
$\bar{D}~D^* + D~\bar{D}^*$ cross section.}}
\label{fig2}
\end{figure}

For $\psi DD,~\psi D_sD_s,~\psi D^* D^*$ and $\psi D_s^* D_s^*$ couplings 
we follow refs.~\cite{mamu98,linko} and make use of the vector meson dominance 
model (VDM). We get 
\beq
g_{\psi DD}=g_{\psi D_sD_s}=g_{\psi D^* D^*}=
g_{\psi D_s^* D_s^*}= 7.64.
\label{vmd1}
\eeq
Using the above SU(4) relations we get
for the other coupling constants: 
\beq
g_{K D_sD^*}=g_{K DD_s^*}=6.6~,
\label{vmd2}
\eeq 
and $g_{K \psi D_sD^*}=g_{K \psi DD_s^*}=50.55$.

The solid line in Figure 2 shows the total cross section of $J/\psi$ 
dissociation by kaons as a 
function of the initial  energy $\sqrt{s}$. The dot-dashed line includes the
contribution for both $K J/\psi \rightarrow D_s {\bar D}^*$ and $\bar K J/\psi 
\rightarrow D^* {\bar D}_s$, while the dotted line includes the
contribution for both $K J/\psi \rightarrow D_s^* \bar D$ and $ \bar K J/\psi 
\rightarrow D {\bar D}_s^*$. In Fig.~2 we also show, for comparison, the
cross section for the process $\pi J/\psi \rightarrow D^* \bar D+{\bar D}^*D$
(dashed line) using the same coupling constants as in ref.~\cite{linko}.
We see that, considering all allowed process for the $J/\psi$ dissociation
by kaons, we get a cross section bigger than the pion-$J/\psi$
dissociation cross section. 

If instead of using the VDM to determine the couplings, we follow 
ref.~\cite{haglin} and use the experimental value of the $\rho\pi\pi$
coupling constant:
\beq
g_{\rho\pi\pi}={g\over2}=6.06,
\eeq
we get: 
\beqa
g_{\psi DD}=g_{\psi D_sD_s}=g_{\psi D^* D^*}&=&
g_{\psi D_s^* D_s^*}= 4.95\nonumber\\
g_{K D_sD^*}=g_{K DD_s^*}&=&4.3~.
\label{grho}
\eeqa
In this
case we get a much smaller cross section, as can be seen by the solid
line in Fig.~3, where we also show (dot-dashed line) the previous result.
\begin{figure}[htb]
\centerline{\epsfig{figure=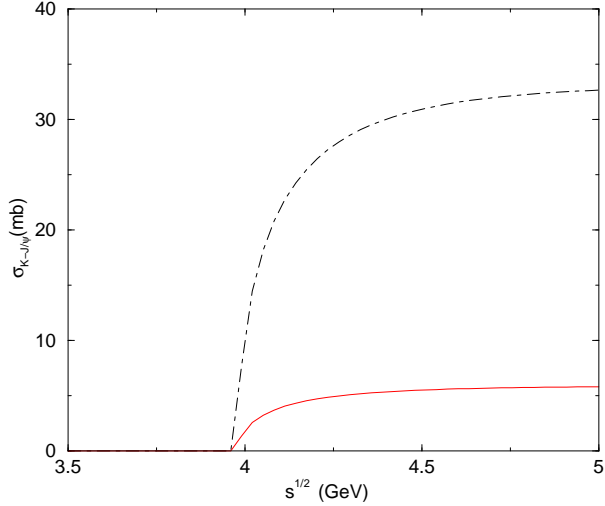,width=8cm}}
\vspace{-1cm}
\protect\caption{\small{Total cross sections of $J/\psi$ 
dissociation by kaons evaluated by using the values for the couplings
given by Eqs.~(\ref{vmd1}) and (\ref{vmd2}) (dot-dashed line) and by
Eq.~(\ref{grho}) (solid line).}}
\label{fig3}
\end{figure}
As can be seen by Fig.~3, the result for the cross section can vary by
almost one order of magnitude, even without considering form factors in the
hadronic vertices \cite{linko,osl}. This gives an idea of how important it is
to have a good estimate of the value of the coupling constants. In a recent
work \cite{haga2}, the $J/\psi-\pi$ and $J/\psi-\rho$ cross sections
were evaluated by using form factors and coupling constants estimated 
using QCD sum rules \cite{nos1,nos2,nos3,nos4}. The results show that
with the appropriate form factors, the total cross section can even fall
for values of $\sqrt{s}$ bigger than 4.5 GeV. In a future work we will
include from factors in the hadronic vertices as well as anomalous parity
interactions \cite{osl}.

In summary, we have studied the cross section of $J/\psi$ dissociation 
by kaons in a meson-exchange model that
includes pseudo-scalar-pseudo-scalar-vector-meson couplings,
three-vector-meson couplings, and four-point couplings.
We find that these cross sections are even bigger than the $J/\psi-\pi\to
\bar{D}D^*+\bar{D}^*D$ dissociation cross section, and
have a very strong dependence
with the values of the coupling constants in the hadronic vertices.
Resulting cross sections can vary between 5 mb and 30 mb for $\sqrt{s}\sim5$ 
GeV, depending on the values of the couplings. 
Since these couplings are not known experimentally, it is very important
to have better estimates for them.

\section*{Acknowledgments}

This work was supported by CNPq and FAPESP.


\vskip5mm
 

\begin{thebibliography}{9}

\bibitem{ma86} T. Matsui and H. Satz, Phys. Lett. {\bf B178}, 416 (1986).
\bibitem{vo99} R. Vogt, Phys. Reports, 310, 197 (1999).
\bibitem{ge99} C. Gerschel and J. Huefner, Ann. Rev. Nucl. Pat. Sci., 
{\bf49}, 255 (1999).

\bibitem{bhp} G. Bhanot and M.E. Peskin, Nucl. Phys. {\bf B156}, 391
              (1979); M.E. Peskin, Nucl. Phys. {\bf B156}, 365  (1979).

\bibitem{kha2} D. Kharzeev and H. Satz, Phys. Lett. {\bf B334}, 155  (1994).

\bibitem{lo} S.H.Lee and Y. Oh, J.Phys. {\bf G28}, 1903 (2002);
             Y. Oh, S. Kim, S.H. Lee, Phys. Rev.  {\bf C65}, 067901 (2002).

\bibitem{wongs} C.-Y. Wong, E.S. Swanson and T. Barnes, Phys. Rev.
                {\bf C62}, 045201 (2000); {\bf C65}, 014903 (2001).

\bibitem{mbq} K. Martins, D. Blaschke and E. Quack, Phys. Rev.
              {\bf C51}, 2723 (1995).

\bibitem{nnmk} F.S. Navarra, M. Nielsen, R.S. Marques de Carvalho and G. Krein,
               Phys. Lett. {\bf B529}, 87 (2002).
%
\bibitem{dlnn} F.O. Dur\~aes, S.H. Lee, F.S. Navarra, M. Nielsen,
               Phys. Lett. {\bf B564}, 97 (2003).
%
\bibitem{dklnn} F.O. Dur\~aes, H. Kim, S.H. Lee, F.S. Navarra, M. Nielsen,
nucl-th/0211092.
%
\bibitem{mamu98} S.G. Matinyan and B. M\"uller, {\sl Phys. Rev.} 
                {\bf C58}, 2994 (1998).


\bibitem{haglin} K.L. Haglin, {\sl Phys. Rev.} {\bf C61}, 031902 (2000).

\bibitem{linko} Z. Lin and C.M. Ko, 
                 {\sl Phys. Rev.} {\bf C62}, 034903 (2000). 

\bibitem{haga} K.L. Haglin and C. Gale, {\sl Phys. Rev.} {\bf C63}, 065201 
(2001).

\bibitem{osl} Y. Oh, T. Song and S.H. Lee, {\sl Phys. Rev.}
              {\bf C63}, 034901 (2001); Y. Oh, T. Song, 
S.H. Lee and C.-Y. Wong, ncl-th/0205065.

%
\bibitem{nnr} F.S. Navarra, M. Nielsen and M.R. Robilotta,
Phys. Rev. {\bf C64}, 021901 (R) (2001).
%
\bibitem{dnn} H.G. Dosch,  F.S. Navarra,  M. Nielsen and M. Rueter, 
Phys. Lett. {\bf B466},  363 (1999). 
%
\bibitem{haga2} K.L. Haglin and C. Gale, hep-ph/0305174.

\bibitem{nos1} F.S. Navarra, M. Nielsen, M.E. Bracco, M. Chiapparini and
C.L. Schat, Phys. Lett.  {\bf B489},  319  (2000). 
%
\bibitem{nos2} F.S. Navarra, M. Nielsen, M.E. Bracco, Phys. Rev.  
{\bf D65},  037502  (2002). 
%
\bibitem{nos3} R.D. Matheus, F.S. Navarra, M. Nielsen and R. Rodrigues da 
Silva, Phys. Lett.  {\bf B541},  265  (2002). 

\bibitem{nos4}  M.E. Bracco, M. Chiapparini, A. Lozea, F.S. Navarra and 
M. Nielsen,  Phys. Lett.  {\bf B521},  1  (2001). 

\end{thebibliography}
\end{document}